\documentclass{sf2a-conf}
\usepackage{graphicx}

\begin{document}
\TitreGlobal{SF2A 2007}
\title{Bar formation and galaxy interactions in MOND}


\author{O. Tiret} \address{Observatoire de Paris, LERMA, 61 Av. de l'Observatoire, F-75014, Paris, France}
\author{F. Combes$^1$}

\runningtitle{Bar formation and galaxy interactions in MOND}

\setcounter{page}{1}

\index{Tiret O.}
\index{Combes F.}

\maketitle

\begin{abstract}
The $\Lambda$CDM model is the most commonly admitted to describe our Universe. In spite of a great success with regard to the large scale structure formation, some problems are still unresolved at galactic scales. Alternative scenarios have to be explored such as modified gravity. We have developed an N-body code able to solve in a self consistent way the galactic dynamics in MOND. The first version of the code consists in solving the modified Poisson equation on a uniform Cartesian grid to derive the gravitational force on each particle. With it, we study the evolution of isolated galaxies, like the bar instability, the angular momentum transfer, etc. Galaxies in MOND are found to form stronger bars, faster than in Newtonian dynamics with dark matter. In a second step, we implement an adaptive mesh refinement technique in the code, allowing to run more contrasted simulations on larger scales, like interacting galaxies. During an interaction, the dynamical friction forces are less important in MOND, and merging times are longer than in DM models. The different morphologies of interacting galaxies in the two models are discussed. All simulations are performed in both frameworks of modified gravity and Newtonian gravity with dark matter with equivalent initial conditions.
\end{abstract}
%

\section{Introduction}

MOND is an alternative theory of gravitation proposed by Milgrom in 1983, to account for galactic dynamics without dark matter. It is based on a modification of the Newtonian gravity below a universal critical acceleration $a_0 \sim 2 \times 10^{-10}$ m s$^{-2}$. In MOND, the Poisson equation is transformed into:
\begin{equation}
\label{eq:mondaqual}
\nabla [ \mu(|\nabla\Phi|/a_0) \nabla\Phi ] = 4 \pi G \rho,
\end{equation}
$\mu(x)$ has the property to be equal to unity for large accelerations (Newtonian regime) and tends to $x$ for the low accelerations (MOND regime). Because of the non-linearity of the MOND gravity, new potential solvers have to be developed for the modified Poisson equation (Brada \& Milgrom 1999, Ciotti et al. 2006, Tiret \& Combes 2007, hereafter TC07). Numerical simulations of spherical dissipationless systems have been carried out by Nipoti et al. 2007. They find that the merging time-scales are significantly longer in MOND compared to Newtonian gravity with dark matter.

We present here N-body simulations of isolated galaxies and interacting galaxies with MOND. Model galaxies have a stellar and gaseous component.

\section{Numerical technique}
\subsection{Uniform grid}

\begin{figure}
	\centering
	\includegraphics[width=5cm]{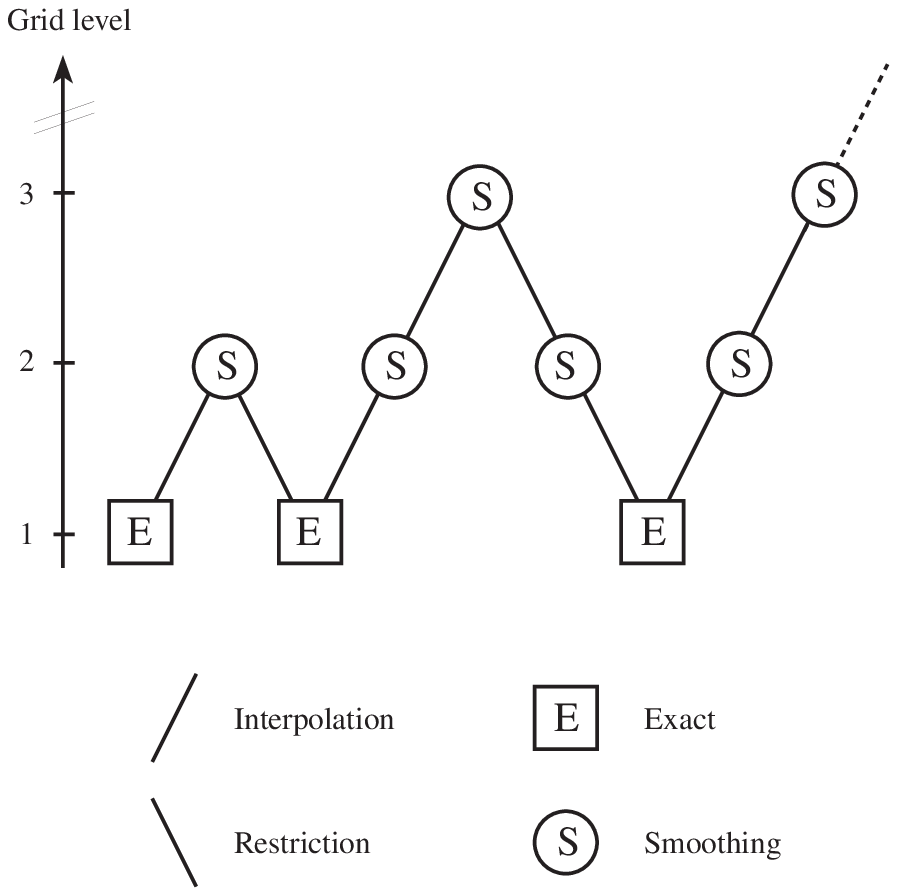}
	\includegraphics[width=5cm]{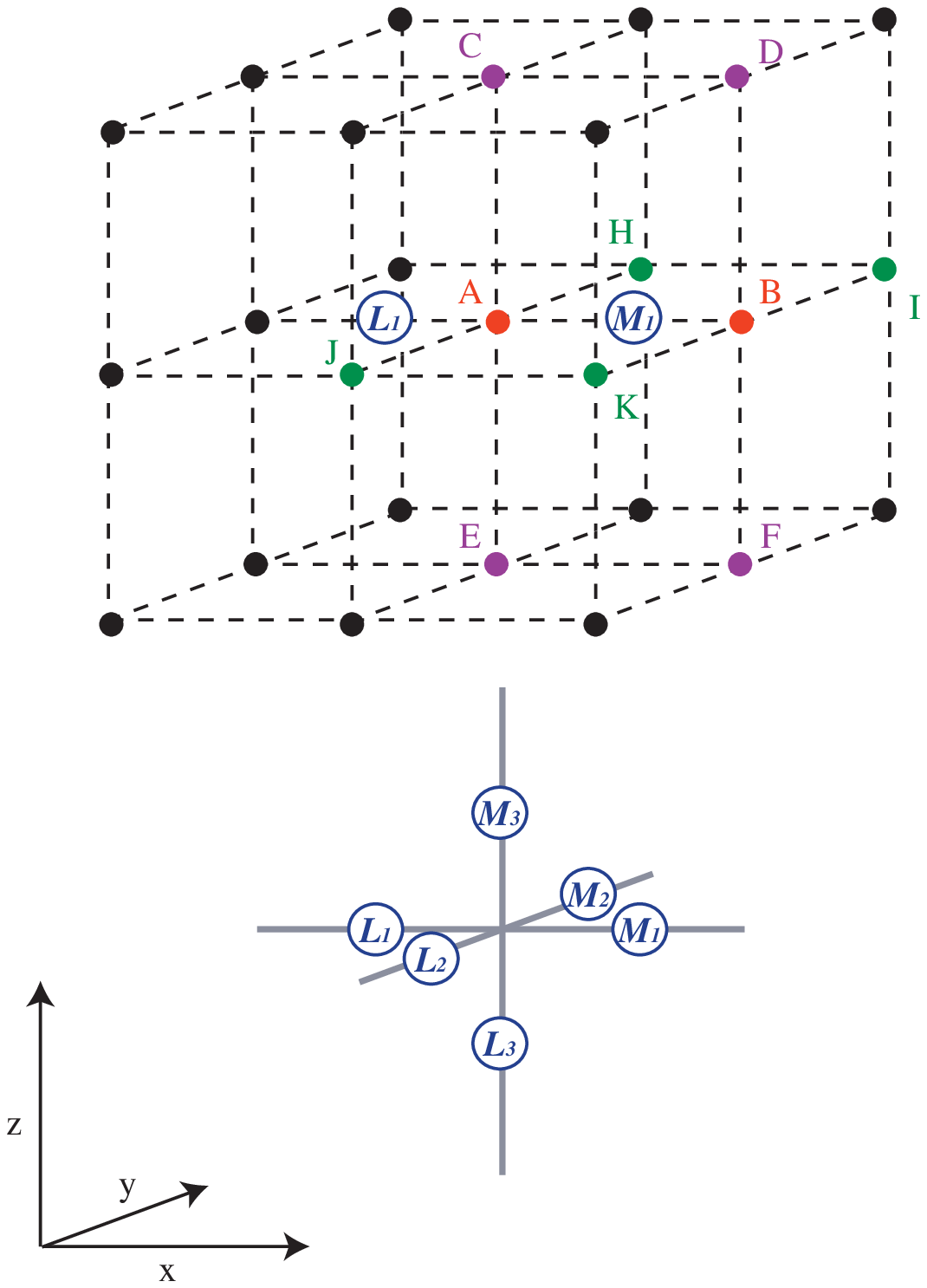}
	\caption{Full multigrid (FMG) algorithm is used to accelerate the convergence in the resolution of the modified Poisson equation.}
	\label{fig:FMG}
\end{figure}

We use a uniform Cartesian grid in three dimensions to study the evolution of an isolated galaxy.
The modified Poisson equation is solved using full multigrid technique (Tiret \& Combes 2007).
It is discretized as follows :
\begin{eqnarray}
\label{eq:monddiscret}
&\ &4\pi G \rho_{i,j,k}= \\ \nonumber
&\ &(\phi_{i+1,j,k}-\phi_{i,j,k})\mu_{M_1}-(\phi_{i,j,k}-\phi_{i-1,j,k})\mu_{L_1} \\ \nonumber
&\ &+(\phi_{i,j+1,k}-\phi_{i,j,k})\mu_{M_2}-(\phi_{i,j,k}-\phi_{i,j-1,k})\mu_{L_2} \\ \nonumber
&\ &+(\phi_{i,j,k+1}-\phi_{i,j,k})\mu_{M_3}-(\phi_{i,j,k}-\phi_{i,j,k-1})\mu_{L_3}) /h^2
\end{eqnarray}
with $\rho_{i,j,k}$ and $\phi_{i,j,k}$ the spatial density and potential discretized on a grid of step $h$, $\mu_{M_l}$, and $\mu_{L_l}$, the value of $\mu(x)$ at points $M_l$ and $L_l$ (Fig. \ref{fig:FMG}).
The gradient component $(\partial /\partial x,\partial /\partial y,\partial /\partial z)$, in $\mu(x)$, are approximated by $({{\phi(B)-\phi(A)}\over h},{{\phi(I)+\phi(H)-\phi(K)-\phi(J)}\over {4h}},{{\phi(C)+\phi(D)-\phi(E)-\phi(F)}\over {4h}})$

Our models evolve in a simulation box of $50$ kpc$^3$, including $256^3$ nodes. The spatial resolution is $400$ pc everywhere.

\subsection{Adaptive refinement resolution}

\begin{figure}
	\centering
	\includegraphics[width=10cm]{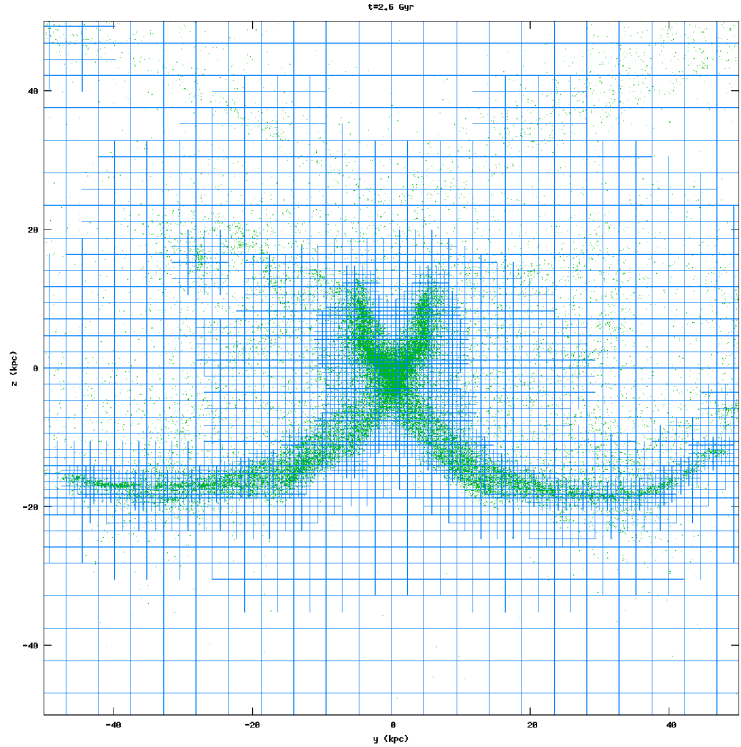}
	\caption{Adaptive mesh structure during a galaxy merger.}
	\label{fig:AMG}
\end{figure}

To simulate interacting galaxies, we now construct an adaptive grid from the mass distribution (\ref{fig:AMG}) and solve the modified Poisson equation (Tiret \& Combes in prep). This allows to use a larger simulation box ($150$ kpc) keeping a good resolution ($\sim 500$ kpc) with reasonable CPU time.

\section{Isolated galaxies}
Galaxies are initially built with a Miyamoto-Nagai disc for the star component, a Toomre disc for the gas and a Plummer sphere for the bulge.
Each galaxy is simulated in the MOND (MOND model) and Newtonian gravity with dark matter (DM model). In the DM model, the DM halo is chosen to obtain the same total rotation curve than in MOND.

\subsection{Bar formation, resonance}
In TC07, we analyze pure stellar discs. Galaxies in MOND are found to form stronger bars and faster than in the DM model. Gas dissipation has now been added using the sticky particle scheme. If the fraction of gas is enough ($\sim 7\%$, in late type galaxies), it makes the galaxy more unstable to bars. As expected, in the new simulations including the gas component, galaxies form bars sooner than without gas, especially in the DM model where galaxies are stabilized by the DM halo.

\begin{figure}
\centering
\includegraphics[width=10. cm]{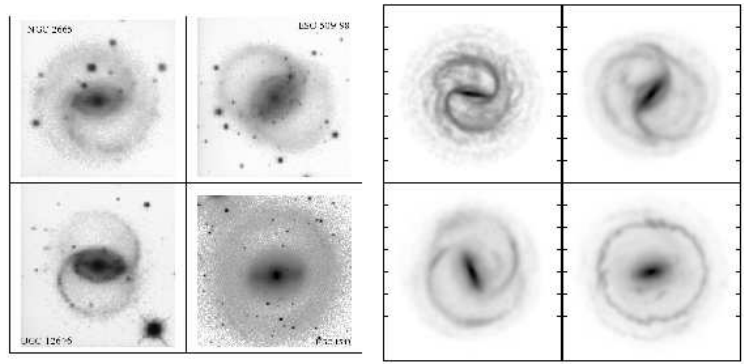}
\caption{Several examples showing the morphological structures of NGC 2665, ESO 509-98, UGC 12646 and NGC 1543 (top panel) compared to simulated galaxies in MOND (bottom panel). Rings and pseudo-rings structures are well reproduced with modified gravity.}
\label{fig:ring}
\end{figure}

In Newtonian dynamic with DM, the bar is slowed down by dynamical friction against DM particles. On the contrary, in MOND, the bar pattern speed remains constant. It implies differences on the resonance locations (like the corotation, ILR, OLR,...). If the bar pattern speed is constant (MOND), the resonance occurs always at the same radius. If the bar slows down (DM), the resonances are shifted outward in the disc. Fig. \ref{fig:ring} shows the apparition of resonant rings and pseudo rings in MOND, corresponding well to the observations.

\subsection{External field effect: EFE}

In a deep MOND regime, the potential of an isolated distribution of finite mass is logarithmic. However the internal potential return to a Keplerian behaviour when an external field is applied. It is why an escape velocity exists also in MOND if the system is embedded in an external potential (Wu et al. 2007).

\begin{figure}
\centering
\includegraphics[width=10. cm]{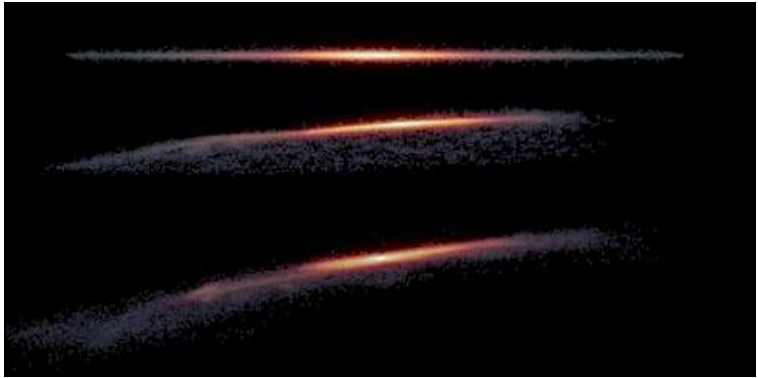}
\caption{Edge on view (t = $0$, $0.5$, $1$ Gyr ) of a galaxy embedded in an acceleration field tilts from $45^\circ$ from the galactic plane.}
\label{fig:efe}
\end{figure}

Let us consider a uniform acceleration field. In Newtonian dynamic, a galaxy will be accelerated uniformly but the evolution of the galaxy in its rest frame is the same as if there was no external field. Due to the non-linearity of the MOND gravity, a galaxy embedded in a uniform acceleration field will have an internal  dynamics which is different from the isolated case. Gravitational torques act on the galaxy and make it precess (Fig. \ref{fig:efe}).

\section{Interaction of galaxies}

With the adaptive multigrid code, we are now able to study interacting galaxies.

\subsection{The antennae}

\begin{figure}
\centering
\includegraphics[width=10. cm]{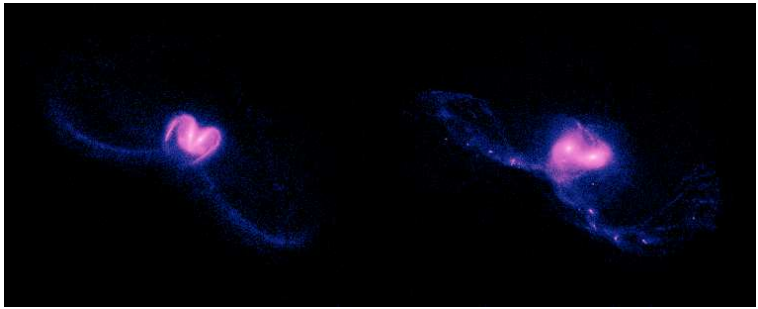}
\caption{Simulations of the Antennae galaxies in the DM model (left) and MOND model(right).}
\label{fig:ant}
\end{figure}

Our first aim was to reproduce the antennae galaxies, by comparing the morphology of this system in the different models, the result is plotted on fig. \ref{fig:ant}.

MOND can form quite extended tidal tails as in the DM model. However, the conditions to obtain a merger are noticeably different. In DM, the halo plays an important role to decrease the distance between the two galaxies by dynamical friction. If the galaxies start on circular orbits, they will decay and merge in $\sim 3$ relative orbits. In MOND, the dynamical friction is efficient only when the two baryonic galaxies overlap each other. Starting from circular orbits, galaxies in MOND could stay bound along a Hubble time without merging. On the other hand, if the orbits are elliptical (with an impact parameter lower than the visible size of the galaxy) merging is possible in a few orbits too.

\subsection{Tidal dwarf}

\begin{figure}
\centering
\includegraphics[width=10. cm]{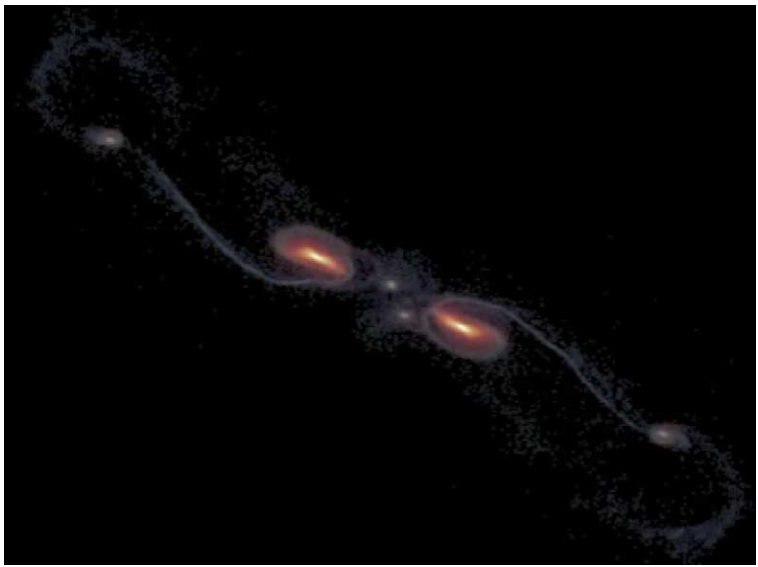}
\caption{Tidal dwarf formation at the tip of the tidal tail in MOND.}
\label{fig:dwarf}
\end{figure}

Several observations of interacting galaxies show the formation of tidal dwarves ($10^8$, $10^9$ $M_\odot$) at the extremity of the tidal arms (ARP105, NGC 7253, ...). In numerical simulations, it is easy to form clumps all along the tidal tails, but this dwarves are difficult to obtain with standard initial conditions. Bournaud et al. (2003) have shown that if the DM halo is extended at least to $10$ times the optical radius, the gravitational potential is strong enough to form dwarf galaxies at the tip of the tail. 
The logarithmic potential of the asymptotic behaviour of MOND plays naturally the role of an extended DM halo. Therefore these tidal dwarves are also obtained in MOND as can be seen on fig. \ref{fig:dwarf}. 

\section{Conclusion}
Numerical simulations with MOND become more and more realistic by taking into account gas dissipation and star formation. With the adaptive mesh refinement code, we are now able to study isolated galaxies as well as interacting galaxies in very short CPU time (several hours of mono-processor).

One particular effect observed in all these simulations is the lack of dynamical friction experienced by the baryonic particles against the DM halo, which does not exist in MOND. For isolated galaxies, the dynamical friction of the DM model makes the bar slow down. One way to increase again the bar pattern speed is to accrete external gas. For interacting galaxies, dynamical friction against the DM halo makes the galaxies merge in less than a Gyr. On the contrary, to obtain a merger in MOND, galaxies require a smaller impact parameter.

Then large differences are expected for the cosmic merger history in MOND and DM model. A given merger will have a longer life-time, even in the appearance of the most perturbed morphology (tidal tails, asymmetries...). To discriminate between the two models, it will be necessary to determine how many mergers a galaxy experienced in its life, still an uncertain parameter.

The fact that MOND implies longer merger times can solve the problem of the existence of compact group of galaxies. The standard CDM scenario predicts that compact groups should have merged a long time ago while too many of them are observed now.

\end{document}